\newcommand{\be}{\begin{equation}}
\newcommand{\ee}{\end{equation}}
\newcommand{\ba}{\begin{array}}
\newcommand{\ea}{\end{array}}
\newcommand{\beqn}{\begin{eqnarray}}
\newcommand{\eeqn}{\end{eqnarray}}

\newcommand{\zero}{\setcounter{equation}{0} \hfill }

\documentclass[12pt]{article}
\usepackage[dvips]{graphicx}

\begin{document}
\title{ Consumer finance data generator - a new approach to 
Credit Scoring technique comparison}
\author{Karol Przanowski\\
Warsaw School of Economics –- SGH \\
Institute of Statistics and Demography \\
Event History Analysis and Multilevel Analysis Unit \\
ul.Madalinskiego 6/8, 
02-513 Warszawa  \\
email: \texttt{kprzan@sgh.waw.pl} \\
url: 
\texttt{www.sgh.waw.pl/zaklady/zahziaw/english/}\\
\texttt{kprzan.w.interia.pl}\\
\and Jolanta Mamczarz\\
Warsaw School of Economics –- SGH \\
University of Warsaw \\
Faculty of Mathematics, Informatics and Mechanics (MIM)
}
\date{}
\maketitle
\begin{abstract}
This paper aims to present a general idea of method comparison of
Credit Scoring techniques. Any scorecard can be made in various
methods based on variable transformations in the logistic regression
model. To make a comparison and come up with the proof that one technique is better than another is a big challenge due to the limited availability of data. 
The same conclusion cannot be guaranteed when using other data from 
another source. The following research challenge can therefore be formulated: how should the comparison be managed in order to get general results that are not biased by particular data? The solution may be in the use of various random data generators. The data generator uses two approaches: transition matrix and
scorings. Here are presented both: results of comparison methods
and the methodology of these comparison techniques creating. Before building a new model the modeler can undertake a comparison exercise
that aims at identifying the best method in the case of the particular data. Here are
presented various measures of predictive model like: Gini, Delta Gini, VIF and Max p-value, emphasizing the multi-criteria problem of a "Good
model". The idea that is being suggested is of particular use in the model building process where there are defined complex criteria trying to cover the important problems of model stability over a period of time, in order to avoid a crisis. Some arguments for choosing Logit or WOE approach as the best scorecard technique are presented.
\end{abstract}
\setcounter{section}{0}
\setcounter{page}{1}
\def\gw{\vskip0.5cm
\centerline{***}
\vskip0.5cm}

\centerline{{\bf Key words:} credit scoring, crisis analysis, banking data generator,}
\centerline{retail portfolio, scorecard building, predictive modeling.}



\section{Introduction \label{intro}}
\zero
\par
Credit Scoring today is applied in various business areas. It especially
has an important usage in the banking sector~\cite{huangimportant}, 
to optimize credit acceptance processes and for the PD models (probability of default) used in Basel II and III for RWA (Risk Weighted Assets) calculations~\cite{baselbis}. 

Their influence on business process has resulted in Credit Scoring becoming a popular and well-known field, yet it remains an area that still requires further development due to the existence of various consultancy companies and corporations, who, because it can be very profitable, often formulate so-called expert statements or methods without having conducted any extensive and fully scientific research. Sometimes this is due to legal constraints that do not allow advance research on particular real data coming from banking processes to be conducted.

Yet, the current crisis demands that researches focus on better predictive modeling, especially with better stability properties in the case of risk over 
time~\cite{stablesupsegment}.

All the above-mentioned arguments suggests the following base questions:

\begin{itemize}
\item Is it possible to conduct Credit Scoring research without any real data?
\item Can a method be formulated that will enable comparison of one technique with another without particular real data, or in other words, can a general data repository for comparisons be created?
\item Can such a general Credit Scoring repository be made available for
all interested parties and will it contain enough particular cases to become GENERAL? 
\end{itemize}

\section{Data used for analysis \label{dane}}
\zero
\par

Two kinds of real data coming from quite different areas (banking and medicine) are used to present the idea of a random data generator as a generalized data for Credit Scoring.

\subsection{Real banking data}

Banking data are taken from one of the Polish banks from the Consumer Finance division. There are $50,000$ rows and $134$ columns. Column names are secured. Target variable represents the typical default event delinquency of more than 60 past due days since the start of the 6 months observation point.

\subsection{Medical real data}

The medical data represents breast cancer survivability 
in USA~\cite{medyczne}. 
The data comes from Surveillance, Epidemiology and End Results repository\footnote{ http://seer.cancer.gov, accessed 30~August~2012.}.
There are $1,343,646$ rows and $40$ columns. Target
function represents either survivability or fatality due to cancer
during the 5 years following diagnosis. The advantage of this data is that there is a large number of rows available, a situation unlike that found in the real banking field.

\subsection{Random data generator}

The Consumer Finance data generator is described by~\cite{karol_gen1}.
The general idea is based on the Markov process with transition matrix. The matrix 
is changing over time due to the impact of one macroeconomic variable. It
results in cyclic risk over time. Every new month of data that is created is based on the score for all credit accounts; cases with greater delinquency have worse scores.
Their shares are connected to particular transition matrix coefficients. Even if the scoring formula for the following months is known, the normal scoring models built in the conventional manner are based on different target functions and can be quite different from the one in the data generator. Despite the simple construction of the data generator, it can be extended and further developed for various portfolios: with small, medium and large risk value (using a different
transition matrix), with small, medium and large periodical property and
different time dependent scoring rules. It is a very flexible way of data creation and the provision of comprehensive information about the process, because not all  the information is secured. All variables and the various form of characteristics that are created can therefore be interpreted. Dataset contains $2,694,377$ rows and $56$ columns.

\section{Steps to follow in scorecard model building}
\zero
\par

For all three kinds of data there are run algorithms of predictive models building. All calculations are made by SAS System\footnote{
SAS Institute~Inc. http://www.sas.com, accessed 30~August~2012.}
based on units: Base SAS, SAS/STAT and SAS/GRAPH.

\begin{itemize}
\item Random samples - data partitioning. Here two datasets are created:
training and validating taken at different times; validating data being taken later. This method - called time sampling - allows to study a model’s stability over time. 
\item Attribute creating - binning. Based on Entropy in order to measure every continuous variable, which is then categorized into an ordinal variable. Some categorical variables are also changed by joining some categories based on similar risk measures. These methods are usually implemented in tree decision techniques.
\item Variable pre-selection - the dropping of insignificant variables. At this stage any information that is based on simple one-dimensional criteria is excluded as they are considered to be variables with little chance of being useful in the next steps. Here predictive powers of single variables and their stability over time are examined. Variables with small powers or those that are significantly unstable are deleted.
\item Multi-factor variable selection - lists of many models. 
In the SAS Logistic procedure a heuristic selection method for continuous variables based on branch and band technique is implemented~\cite{branch}. It is an extremely useful method to produce many models, namely 700 models as the best 100 models with 6-variables, 7-, ... and 12-variables. 
\item Model assessment. There is not any single and unique good model criterion. Instead, a selection is employed, such as: predictive power:
 ($AR$ in other words Gini~\cite{crookbook}), stability: $AR_{diff}$ - delta Gini (relative difference between predictive powers on training and validating datasets), collinearity measures: $MAX_{VIF}$ - maximal variance inflation factor, 
$MAX_{Pearson}$ - maximal Pearson correlation coefficient on pairs of variables and $MAX_{Con Index}$ - maximal condition index and also significant measures: $MAX_{ProbChiSquare}$ - maximal p-value for variables in the model.
\end{itemize}

\section{Different variable coding and selection \label{kod}}
\zero
\par

A scoring model, though based on the same set of variables, can be estimated in logistic regression on various methods dependent on the coding.

The first way, called REG, is a model without any variable transformation.
In this case the missing imputation step, which is certainly not
trivial and can be quite important, is necessary but the REG method is considered here for an additional scale or mirror, so the simplest missing imputation method - imputation by the mean - can be employed.

The second way called LOG is based on logit transformation: for every
attribute (after binning) its logit is calculated. The transformed variable becomes partially constant and discrete (quasi-continuous). This way is useful, because the missing imputation is not required. The missing value can be assigned to a separate attribute or combined with other values dependent on the binning criteria. Moreover, this method treats qualitative and quantitative variables in the same way;
at the end all variables are binned and transformed into a logit structure. This is a similar WOE approach used in SAS Credit Scoring Solution~\cite{sasbook}.

The third way - GRP, is connected to the binary coding called reference or dummy, see table~\ref{dummy}. The reference level is set at the attribute with the lowest risk. Any other solutions where the reference level is, for example, set at the most representative attribute, with the greatest share or other can be considered, though this is a topic for further research. Dummy coding produces a large number of binary variables and it is not easy to run the heuristic branch and
band variable selection method because the time of calculation is increasing
to infinity. It is a typical case of the familiar NP-complete problem. Moreover the company Score Plus~\cite{scoreplus} rightly suggests to run the selection method based on better coding called ordinal or nested, see table~\ref{nesteddes},~\ref{nestedasc} and~\ref{nestedmon}. 
In the case of the last mentioned coding method, all betas in the model with one variable have the same sign, but this experimental fact requires formal proof.

In the cases of REG and LOG one single beta is estimated for every variable in the model. For the GRP method every beta is estimated separately for every attribute, so in that case the number of parameters in the model is about 6 times greater (if we assume 7 attributes per variable). Another good research topic would be to take the following into consideration: diagnostic research of GRP models, their correctness of estimation, minimal sample size and powers of statistical tests.
Intuition suggests that care should be taken here because models can be overestimated.

In the case of GRP, due to a lack of variable heuristic selection
all variable combinations resulting from the REG and LOG methods are taken. All these combinations taken together are estimated by the GRP method.

In practice it is often the case that by using the GRP method some attributes are not significant, but the whole variable can be significant, especially by "TYPE~3" tests. Yet, a single attribute remains insignificant. It is not advisable to retain that attribute in the final model. What is needed is a new sub-method to eliminate insignificant attributes when using the GRP way. Without that step all
results of GRP do not provide good models to become a serious competitor to LOG. In order to be so a solution for the elimination of insignificant
attributes called attribute adjustment should be devised. 
Here are chosen two simple algorithms: backward and stepwise, all available in SAS Logistic procedure.

The model can be estimated based on dummy coding or nested. Therefore finally 12 attribute adjustments methods are created, see table~\ref{korekty}.

\begin{table}
\begin{center}
\caption{Example of scorecard model.}
\label{karta}

{\scriptsize

\begin{tabular}{ c | c | c}
\hline
Variable & Condition (attribute) & Partial score \\
\hline
Age &  

\begin{tabular}{ c }
$\le$ 20 \\
$\le$ 35 \\
$\le$ 60 
\end{tabular} &

\begin{tabular}{ c }
 10 \\
 20 \\
 40
\end{tabular}

\\ \hline
Income & 
\begin{tabular}{ c }
$\le$ 1500 \\
$\le$ 3500 \\
$\le$ 6000 
\end{tabular} &

\begin{tabular}{ c }
 15 \\
 26 \\
 49
\end{tabular}

\\ \hline
\end{tabular}
}
\end{center}
\end{table}

All models with exclusion REG are scorecard models, see table~\ref{karta}.

\begin{table}
\begin{center}
\caption{Reference coding - dummy.}
\label{dummy}

{\scriptsize

\begin{tabular}{ c c c c }
Group number & Variable1 & Variable2 & Variable3 \\
\hline
1 & 1 & 0 & 0 \\
2 & 0 & 1 & 0 \\
3 & 0 & 0 & 1 \\
4 & 0 & 0 & 0 
\end{tabular}
}
\end{center}
\end{table}

\begin{table}
\begin{center}
\caption{Cumulative descending coding - nested descending (ordinal).}
\label{nesteddes}

{\scriptsize

\begin{tabular}{ c c c c }
Group number & Variable1 & Variable2 & Variable3 \\
\hline
1 & 0 & 0 & 0 \\
2 & 1 & 0 & 0 \\
3 & 1 & 1 & 0 \\
4 & 1 & 1 & 1 
\end{tabular}
}
\end{center}
\centerline{
{\scriptsize
Source: SAS Institute Inc. 2002-2010. SAS/STAT 9.2: Proc Logistic - 
User's Guide, Other Parameterizations.}}
\end{table}

\begin{table}
\begin{center}
\caption{Cumulative ascending coding - nested ascending.}
\label{nestedasc}

{\scriptsize

\begin{tabular}{ c c c c }
Group number & Variable1 & Variable2 & Variable3 \\
\hline
1 & 1 & 1 & 1 \\
2 & 0 & 1 & 1 \\
3 & 0 & 0 & 1 \\
4 & 0 & 0 & 0 
\end{tabular}
}
\end{center}
\end{table}

\begin{table}
\begin{center}
\caption{Cumulative monotonic coding - nested monotonic.}
\label{nestedmon}

{\scriptsize

\begin{tabular}{ c c c c }
Group number & Variable1 & Variable2 & Variable3 \\
\hline
1 & 1 & 1 & 1 \\
2 & 1 & 1 & 0 \\
3 & 1 & 0 & 0 \\
4 & 0 & 0 & 0 
\end{tabular}
}
\end{center}
\end{table}

\begin{table}
\begin{center}
\caption{Attribute adjustments for GRP models}
\label{korekty}

{\scriptsize

\begin{tabular}{ c c c c }
Method name & Estimation & Selection & Coding \\
\hline
NBA & nested  & backward & ascending nested \\
NBD & nested  & backward & descending nested \\
NBM & nested  & backward & monotonic nested \\
NSA & nested  & stepwise & ascending nested \\
NSD & nested  & stepwise & descending nested \\
NSM & nested  & stepwise & monotonic nested \\
DBA & dummy   & backward & ascending nested \\
DBD & dummy   & backward & descending nested \\
DBM & dummy   & backward & monotonic nested \\
DSA & dummy   & stepwise & ascending nested \\
DSD & dummy   & stepwise & descending nested \\
DSM & dummy   & stepwise & monotonic nested 
\end{tabular}
}
\end{center}
\end{table}

\section{Results \label{wyniki}}
\zero
\par

For every kind of data: sample datasets training, validating and variable pre-selections are created, see table~\ref{partition}.

In the next step 700 models for REG and LOG are calculated separately. Then $1,400$ models are estimated by GRP method. Every GRP model then is adjusted by all 12 methods. To summarize about $19,600$ models for every kind of data are created and estimated, so in total about $58,800$ models. Such a large number of models with their various criteria statistics creates the possibility to study distributions of these criteria and to make a thorough comparison based on distribution properties.

All calculations are made on a simple Laptop Core Duo 1,67GHz and take
about 2 months without interruptions to complete.

\begin{table}
\begin{center}
\caption{Sample sizes}
\label{partition}

{\scriptsize

\begin{tabular}{ c c c c }
Data source & Training & Validating & Number of chosen variables \\
\hline
Banking & 27~325 & 12~435 & 60 \\
Medical & 29~893 & 17~056 & 23 \\
Random & 66~998 & 38~199 & 33 
\end{tabular}
}
\end{center}
\end{table}

\begin{figure}
\caption{One–dimensional distributions - prediction.}
\label{jedno_razem_ar_valid}
\vskip0.5cm
\begin{center}
\includegraphics[angle=-90, width=0.9\textwidth]{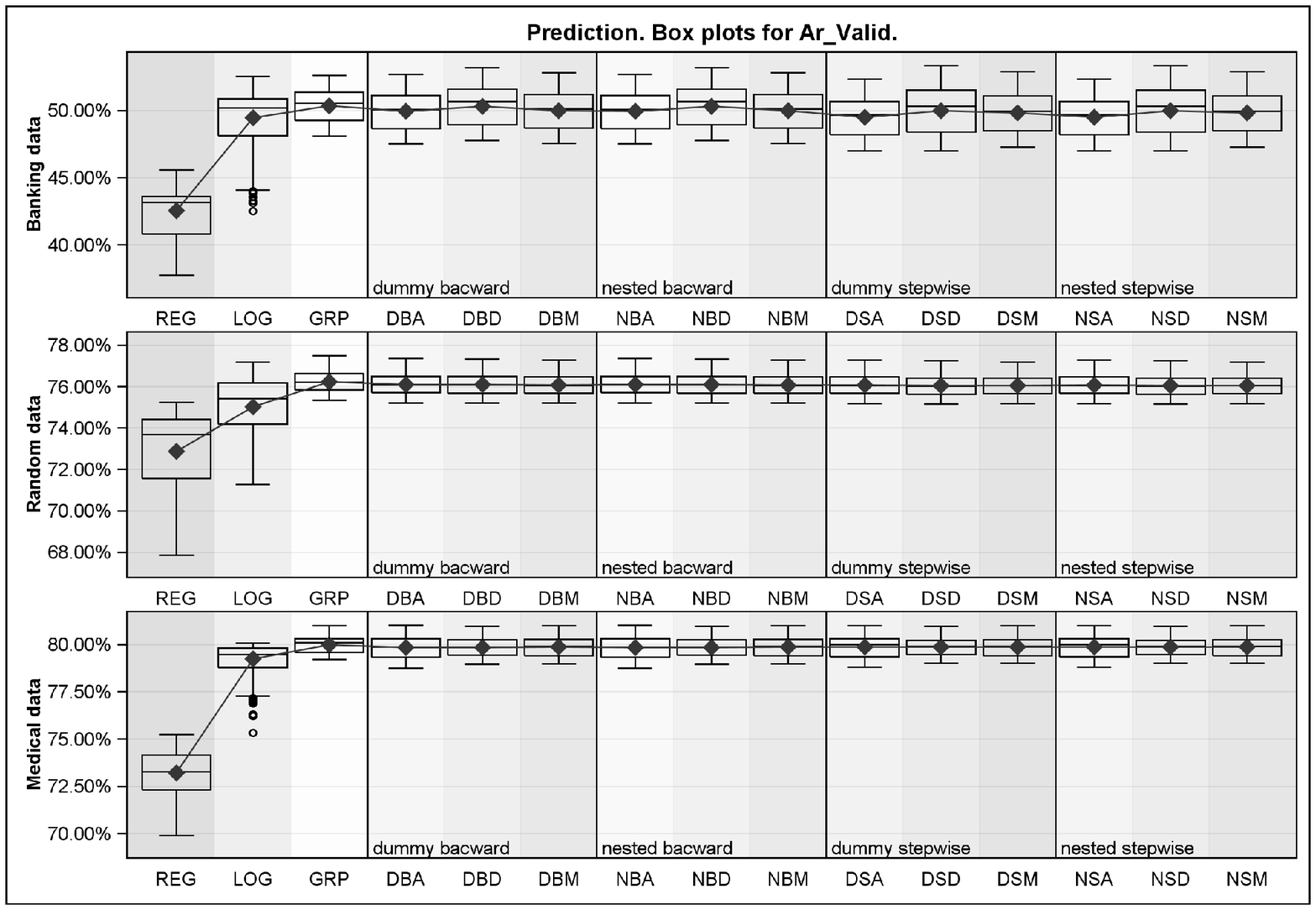}
\end{center}
\end{figure}

\begin{figure}
\caption{One–dimensional distributions - stability.}
\label{jedno_razem_ar_diff}
\vskip0.5cm
\begin{center}
\includegraphics[angle=-90, width=0.9\textwidth]{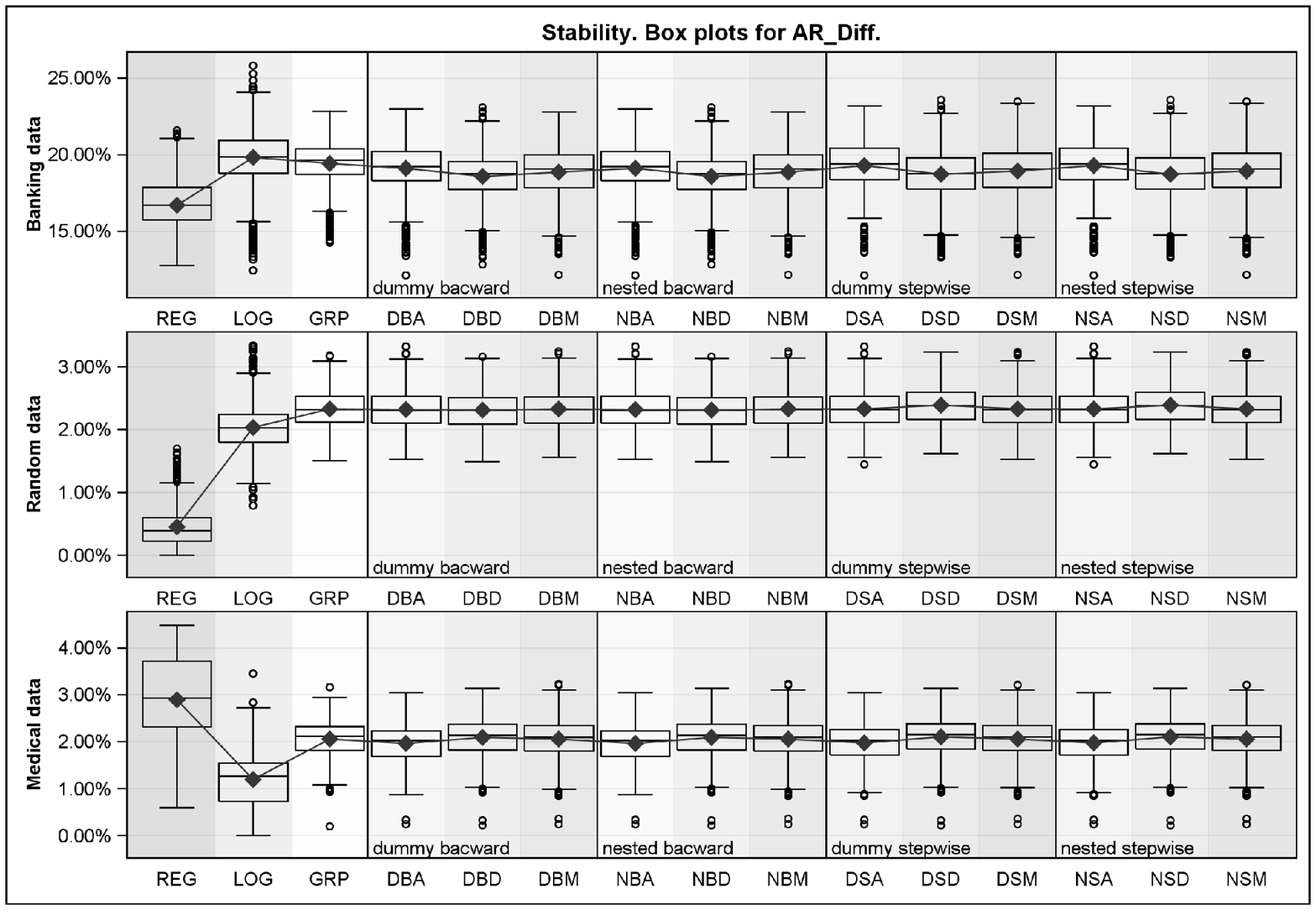}
\end{center}
\end{figure}

\begin{figure}
\caption{One–dimensional distributions - collinearity.}
\label{jedno_razem_vif}
\vskip0.5cm
\begin{center}
\includegraphics[angle=-90, width=0.9\textwidth]{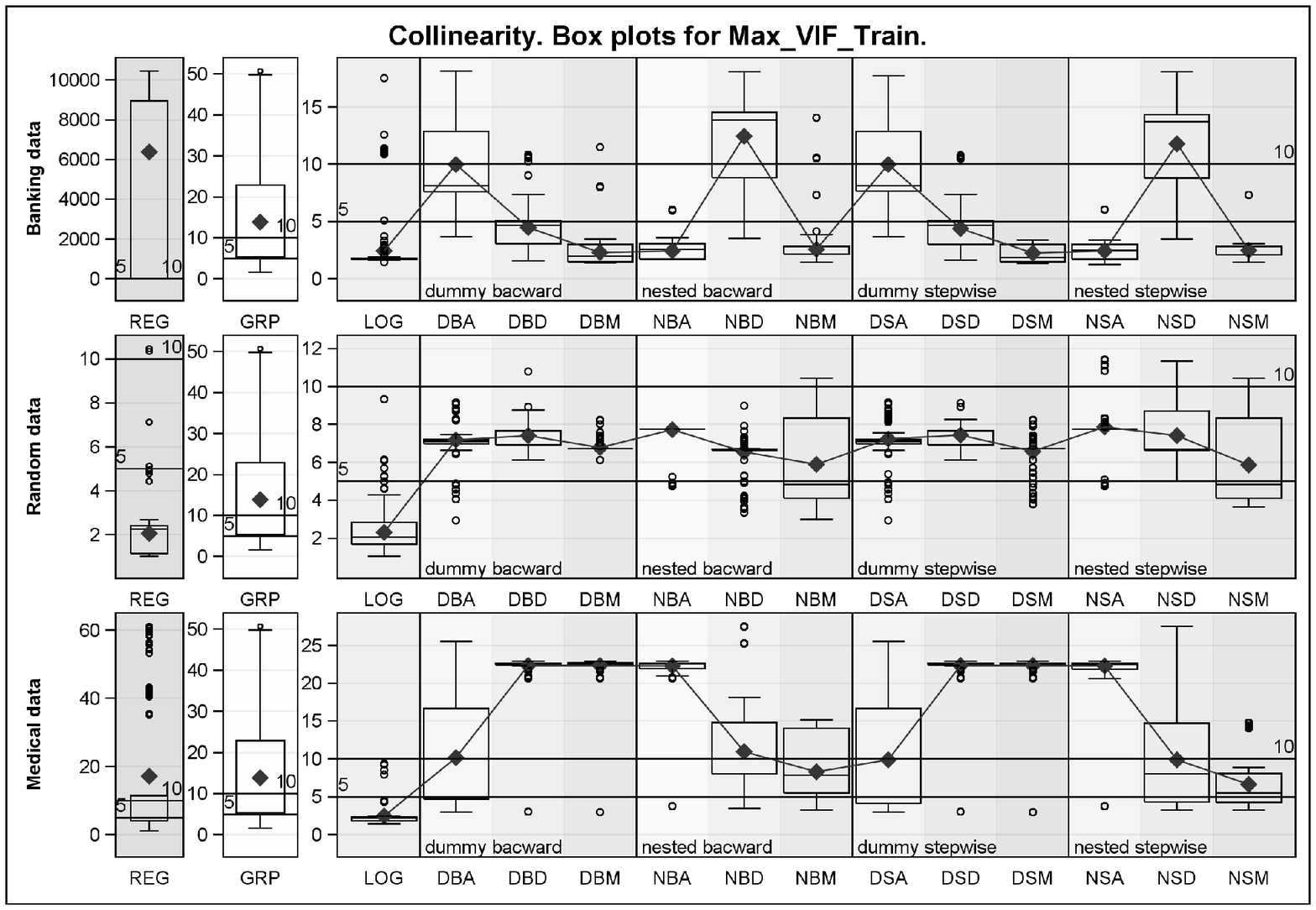}
\end{center}
\end{figure}

\begin{figure}
\caption{Multi–dimensional approach. Stability and prediction with the same weights.}
\label{ideal_razem_stabpredykt}
\vskip0.5cm
\begin{center}
\includegraphics[angle=-90, width=0.8\textwidth]{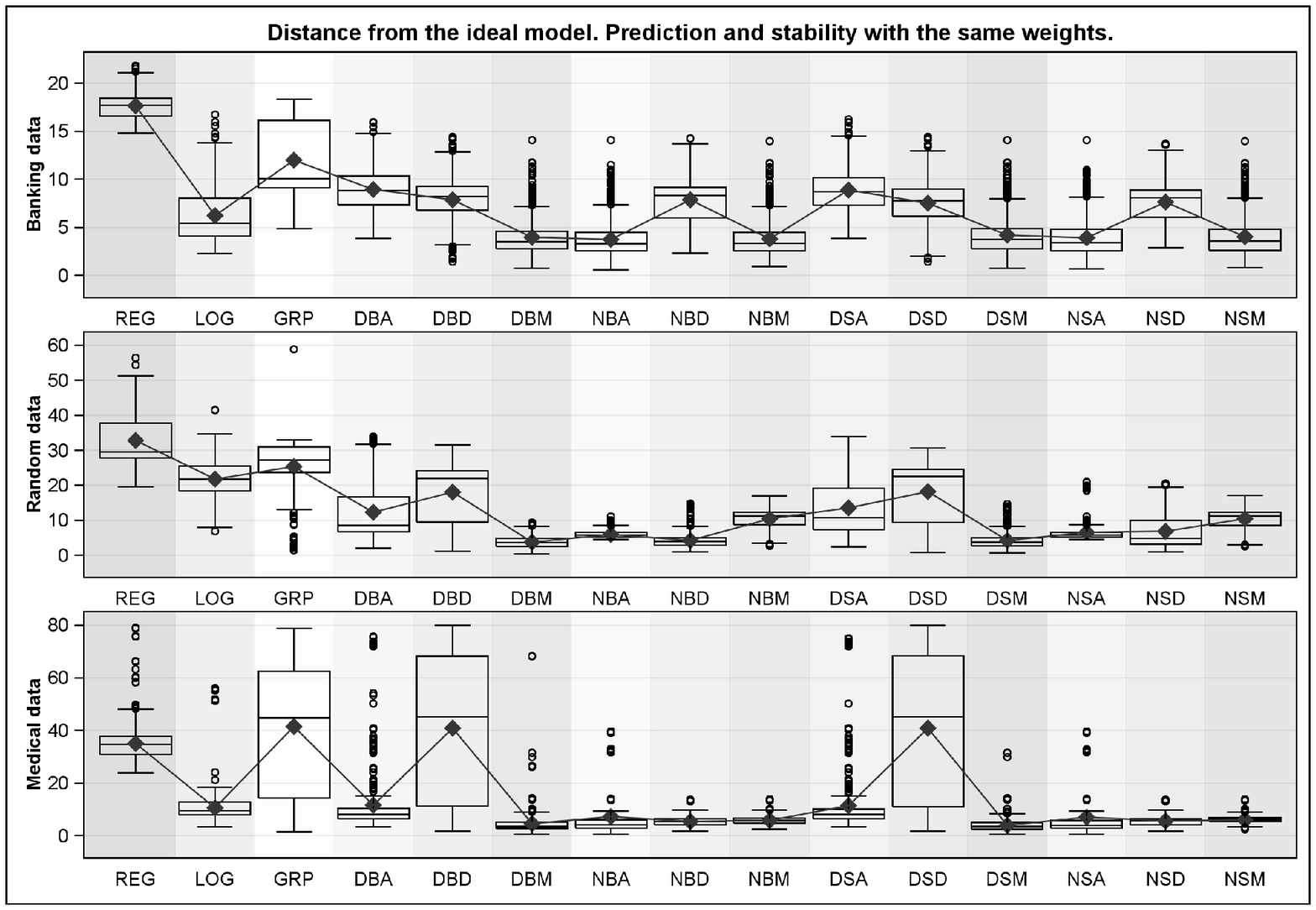}
\end{center}
\end{figure}

\begin{figure}
\caption{Multi–dimensional approach. Stability with greater weight than Prediction.}
\label{ideal_razem_stab}
\vskip0.5cm
\begin{center}
\includegraphics[angle=-90, width=0.8\textwidth]{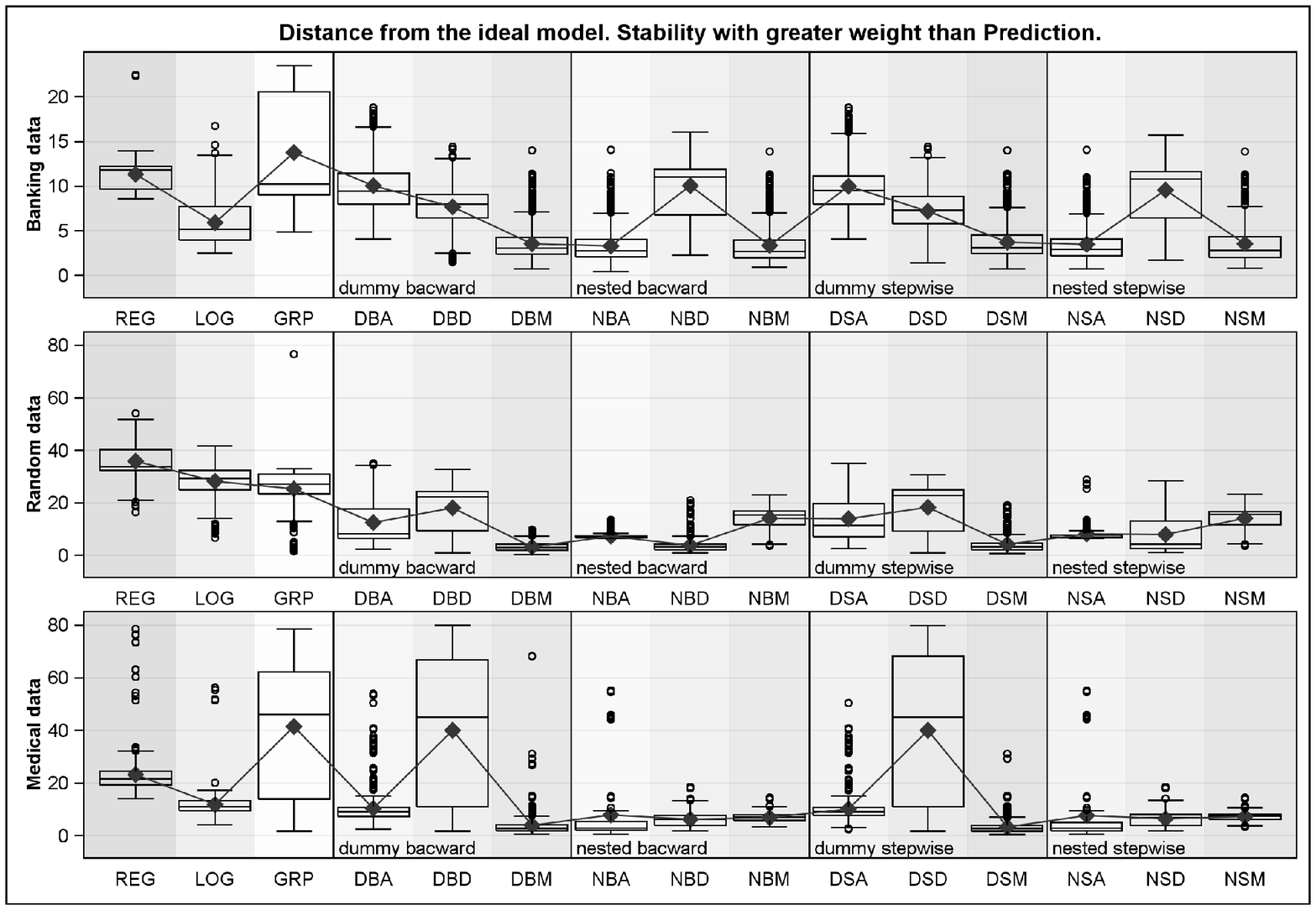}
\end{center}
\end{figure}

\begin{figure}
\caption{Multi–dimensional approach. Prediction with greater weight than stability.}
\label{ideal_razem_predykt}
\vskip0.5cm
\begin{center}
\includegraphics[angle=-90, width=0.8\textwidth]{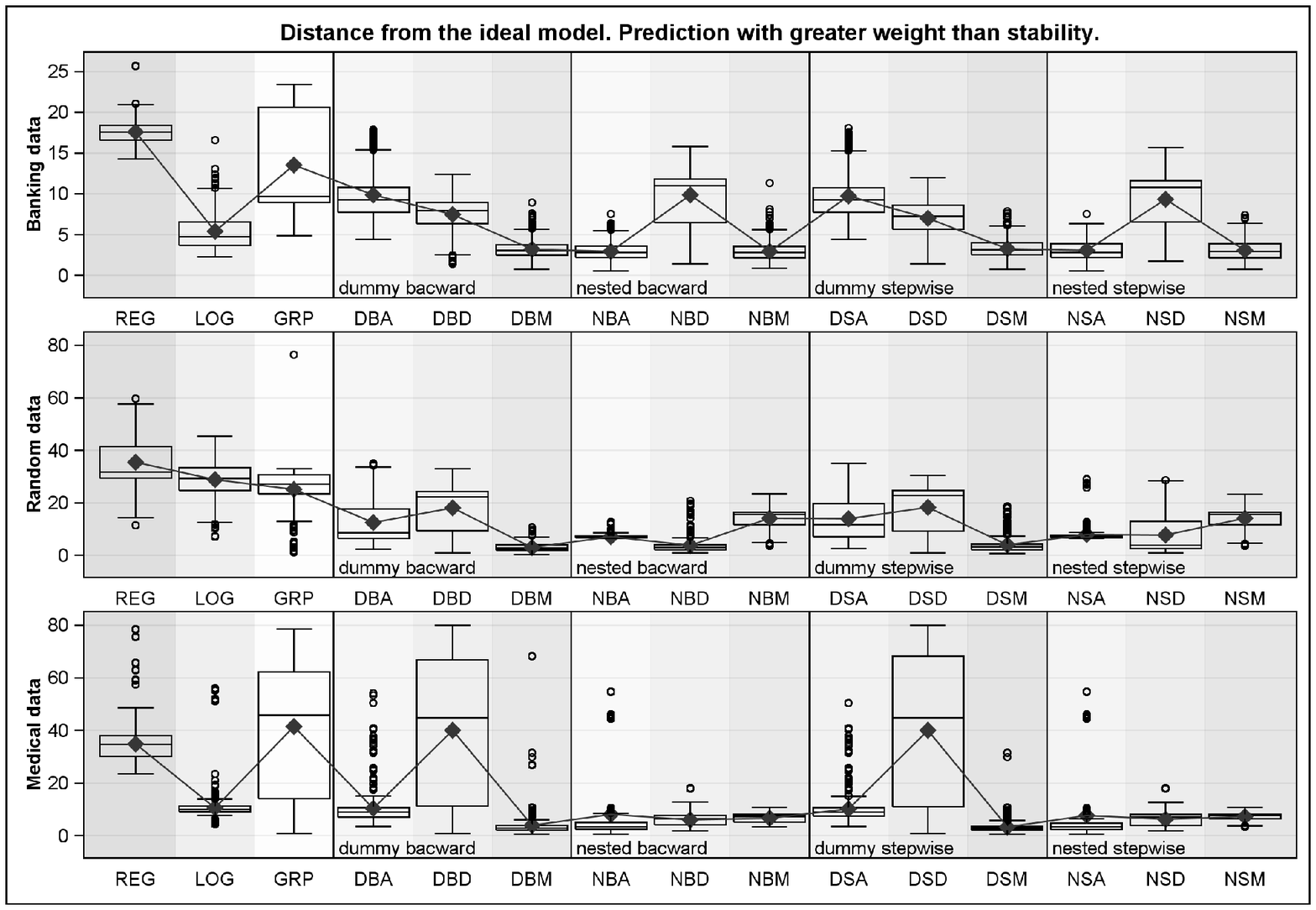}
\end{center}
\end{figure}


\section{Interpretation}
\zero
\par

15 predictive modeling techniques: REG, LOG, GRP and 12 attribute adjustments are calculated and compared.
For every technique mentioned
above and in order to avoid scale problem 700
best models are initially selected. These are based on 
$AR_{Valid}$, e.g. predictive power (Gini statistic) on validating dataset.

In figures~\ref{jedno_razem_ar_valid}, \ref{jedno_razem_ar_diff} and 
\ref{jedno_razem_vif} one-dimensional distributions of the few model criteria: prediction, stability and collinearity are presented. The main differences for prediction using $AR_{Valid}$ can be indicated for models REG, LOG and GRP. All GRP adjustments have similar results.
The same conclusion is true in the case of stability using $AR_{diff}$.
When using collinearity there are significant differences. GRP adjustments strongly improve $MAX_{VIF}$
and for LOG models almost all values concentrate around an acceptable level.

A one-dimensional approach is unable to identify the best scoring techniques in the correct way, because even if one model has the best prediction, it can also have the worst stability, so rather ought to be excluded from the list of suitable candidates.
The better approach is to analyze the multi-dimensional criterion, where all model statistics are taken together and where the distance from the ideal model is defined. The ideal model is the "crystal ball": the highest prediction (100\%), null collinearity and null instability.
It the practice not all criteria have the same weights, but it is not a trivial problem to define the proper priorities. In figures~\ref{ideal_razem_stabpredykt} , \ref{ideal_razem_stab} and \ref{ideal_razem_predykt} three cases with different relations between weights for prediction and stability: equality, minority and majority are presented. The lower note means a better model; one that is closer to ideal model. This manner of data presentation gives quite
interesting results. REG models significantly lie outside the ideal model
for every type of data. The GRP has too large a variance and also is not close to the ideal model. LOG models have desirable notes, but consistently fail to have the lowest note: the minimal distance to the ideal model. Some GRP adjustments have the best properties, especially models estimated by nested coding. Furthermore, all adjustments with monotonic coding are concentrated around very good levels, almost always the minimal distance from the ideal model.

From amongst all the adjustments methods NBM (nested, backward,
monotonic nested) is the one that ought to be highlighted as a good method based on the results presented and with the added bonus of simple implementation and time of calculation. So, in conclusion, only two methods are chosen for further analysis: LOG and NBM.

\section{Final comparison: LOG contra NBM}
\zero
\par

\begin{figure}
\caption{Scatter plots: comparison for LOG and NBM methods. Banking data.}
\label{logvsnbm_ban}
\vskip0.5cm
\begin{center}
\includegraphics[angle=-90, width=0.7\textwidth]{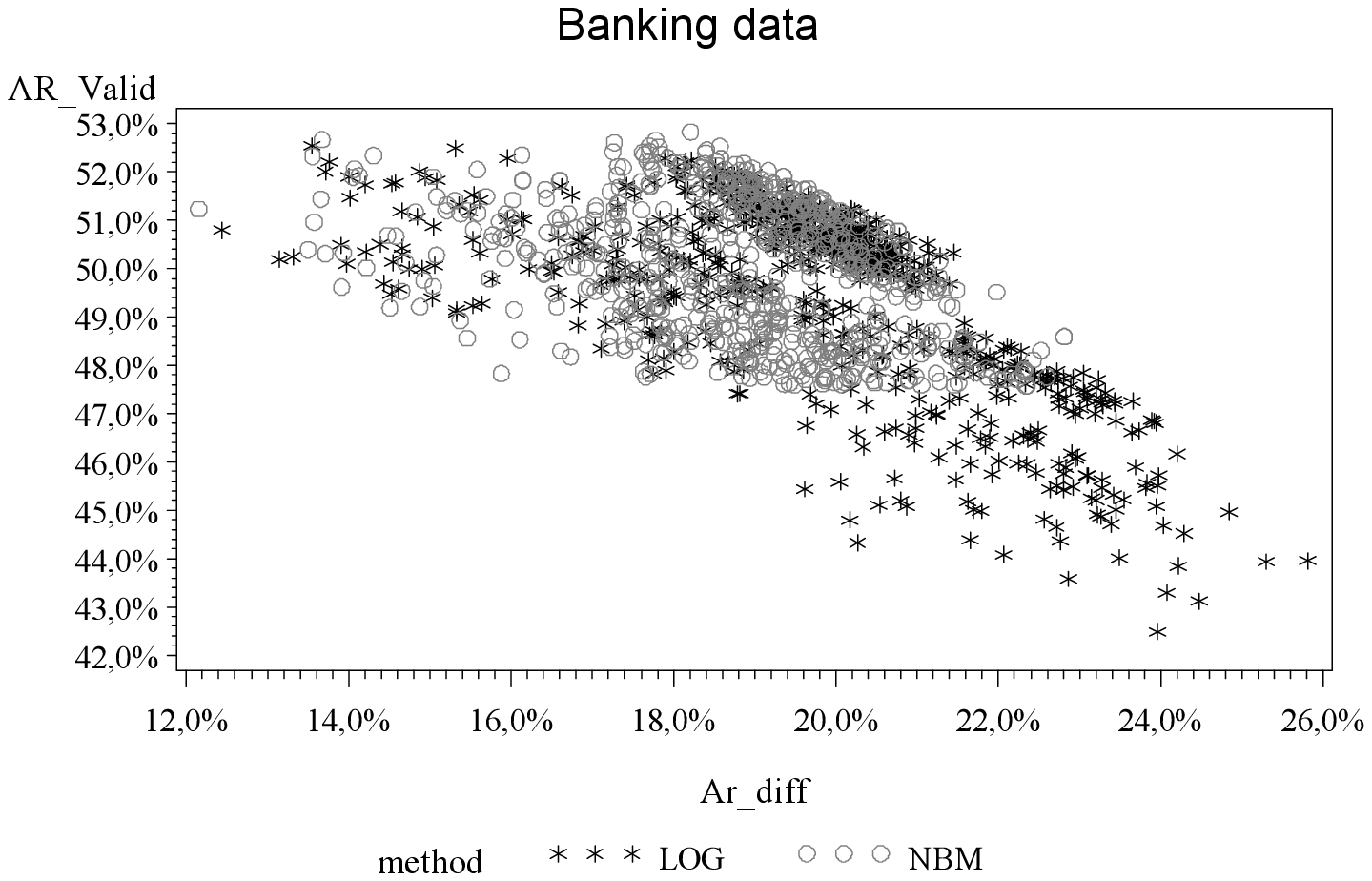}
\end{center}
\end{figure}

\begin{figure}
\caption{Scatter plots: comparison for LOG and NBM methods. Random data.}
\label{logvsnbm_ran}
\vskip0.5cm
\begin{center}
\includegraphics[angle=-90, width=0.7\textwidth]{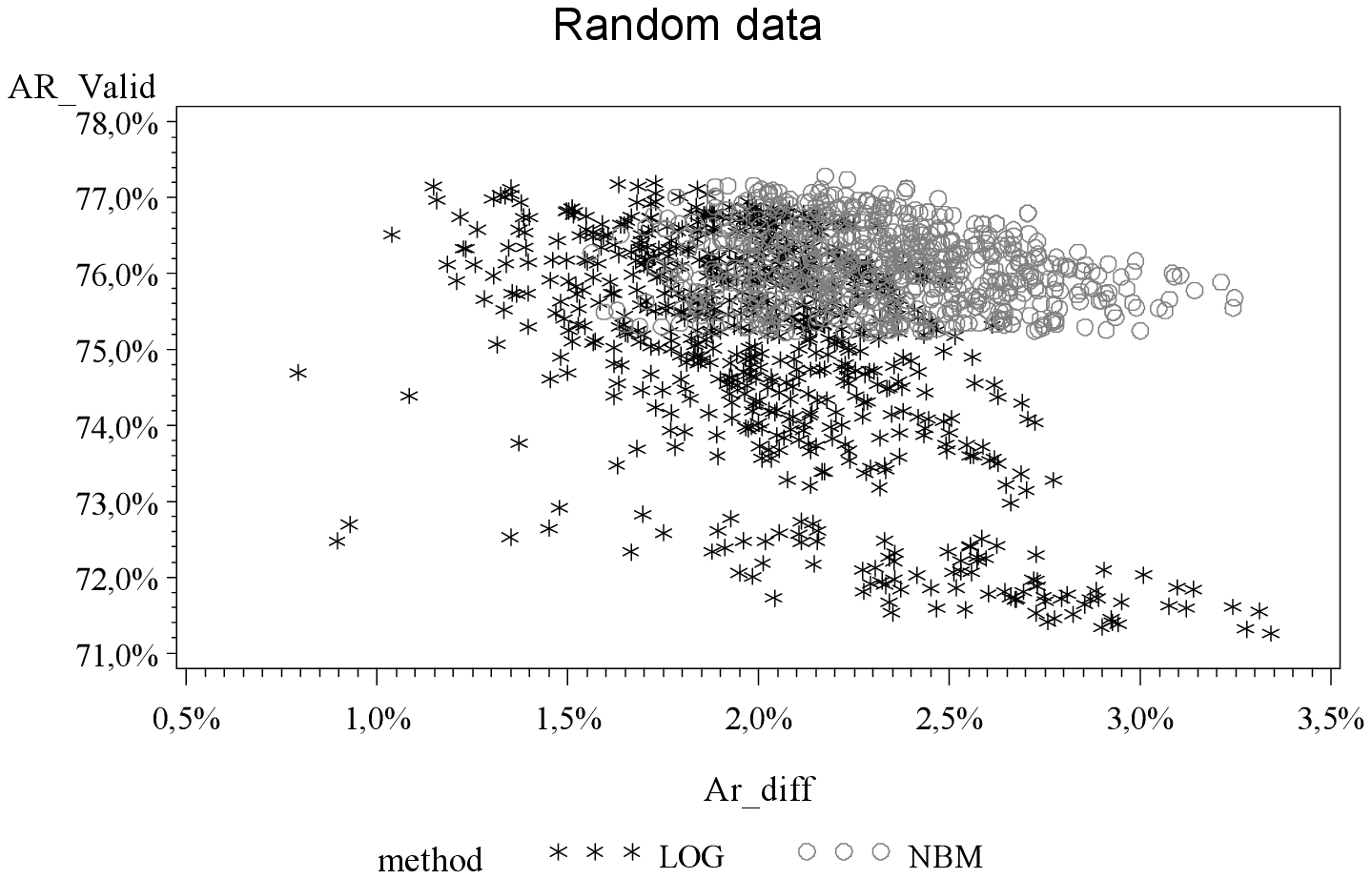}
\end{center}
\end{figure}

\begin{figure}
\caption{Scatter plots: comparison for LOG and NBM methods. Medical data.}
\label{logvsnbm_med}
\vskip0.5cm
\begin{center}
\includegraphics[angle=-90, width=0.7\textwidth]{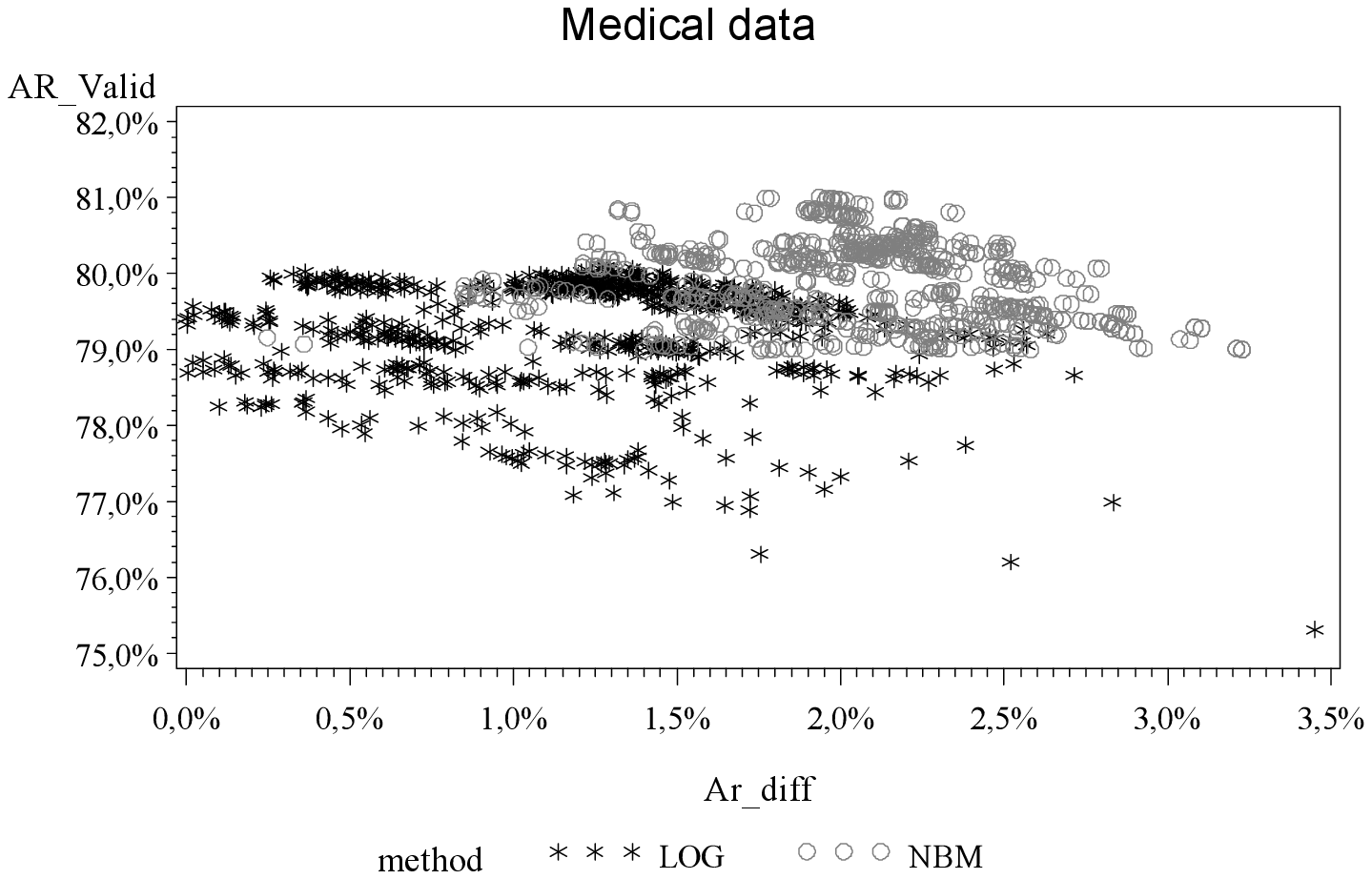}
\end{center}
\end{figure}

Based on many 3D analysis, which cannot be presented in this paper, only two of the most important criteria to identify significant differences between LOG and NBM methods are chosen. Only prediction $AR_{valid}$ and stability $AR_{diff}$ are required to present the final comparison.
In figures~\ref{logvsnbm_ban}, \ref{logvsnbm_ran} and \ref{logvsnbm_med} 
scatter plots of these two statistics for three kinds of data are presented. Here real data from modeling process without
any scaling are presented. It can be indicated that the LOG method (represented by stars on the figure) provides slightly more stable models than NBM (represented by gray circles) and with slightly lower predictive powers than NBM. Because the difference is not very marked and almost always can be found in models with similar properties when using both methods it is suggested that the simplest method, LOG, is used. On the other hand, from these two criteria a more conservative approach is to select models with better stability than greater prediction. So, finally, after various analyses among 15 scoring techniques the LOG method is the simplest and the best method in order to build good models where, for example, the modeler does not have enough time. In other cases it is suggested to always make a serious analysis of all known and available scoring techniques because the best method is a spectrum of methods.

\section{Conclusion \label{podsum}}
\zero
\par
In spite of the three different kinds of data: banking, medicine and random all the comparison results of the various scoring techniques seem to be in convergence and give the same conclusions. In other words the conclusion can be formulated that the research method for scoring technique comparison presented in the paper is independent from data and is not biased by particular data structures.
This is a very profitable statement, prompts further research and gives the possibility to focus on one more available data type: random data. Moreover, the comparison technique which is presented can be always updated for new data. The analyst can always, before building a new model, run the technique presented here in order to see the results directly coming from his data, even they prefer a method based on their own experience. The one disadvantage is the time of calculation. This argument suggests starting many analyses on random data to begin with, because they are always available and can be published without any special restrictions. The random data can, of course, be created in various ways, always be improved upon or altered in order to get better and more general conclusions.

It would now seem possible to answer the main question about the possibility of research in Credit Scoring without real data. Even if the results presented for the three kinds of data have some small differences, the general message is that it is possible to create a General Credit Scoring Data Repository based on some random generators.



\newcommand{\byauthors}[1]{#1 }
\newcommand{\journal}[1]{ #1 }
\newcommand{\reftitle}[1]{{\it #1} }
\newcommand{\volumin}[1]{{\bf #1} }
\newcommand{\eref}{.}
\newcommand{\refyear}[1]{(#1), }







\end{document}